\documentclass[conference]{IEEEtran}

\usepackage{array}
\usepackage{algorithm}
\usepackage{algorithmic}
\usepackage{amsmath}
\usepackage{amssymb}
\usepackage{cite}
\usepackage{epsfig}
\usepackage{float}
\usepackage{graphicx}
\usepackage{mdwmath}
\usepackage{mdwtab}
\usepackage[tight,footnotesize]{subfigure}

\begin{document}

\title{An Algorithm of Parking Planning for\\Smart Parking System}

\author{\IEEEauthorblockN{Xuejian Zhao}
\IEEEauthorblockA{Wuhan University\\
Hubei, China\\
Email: xuejian\_zhao@sina.com}
\and
\IEEEauthorblockN{Kui Zhao}
\IEEEauthorblockA{Zhejiang University\\
Zhejiang, China\\
Email: zhaokui@zju.edu.cn}
\and
\IEEEauthorblockN{Feng Hai}
\IEEEauthorblockA{Wuhan University\\
Hubei, China\\
Telephone: (+86) 027-68753031}
}
\maketitle

\begin{abstract}
There are so many vehicles in the world and the number of vehicles is increasing 
rapidly. To alleviate the parking problems caused by that, 
the smart parking system has been developed. The parking 
planning is one of the most important parts of it. An effective 
parking planning strategy makes the better use of parking resources possible. 
In this paper, we present a feasible method to do parking planning. 
We transform the parking planning problem into a kind of linear assignment problem. 
We take vehicles as jobs and parking spaces as agents. 
We take distances between vehicles and parking spaces as 
costs for agents doing jobs. Then we design an algorithm for this particular assignment problem
and solve the parking planning problem. The method 
proposed can give timely and efficient guide information to vehicles for 
a real time smart parking system. Finally, we show the 
effectiveness of the method with experiments over some data, 
which can simulate the situation of doing parking planning in the real world.
\end{abstract}


\section{Introduction}
There are so many vehicles in the world, especially in China. 
According to the annual report presented by J. Wang
\cite{WangJunxiu2013AnnualReport}, there are almost one 
billion private cars in China till 2013. In other words,  
there are 20 private cars per 100 families. By contrast, 
the parking resources are too limited. Taking Beijing as an 
example, in 2013, there are 4.075 million private cars, but the number of 
parking spaces is only about 1.622 million
\footnote{http://www.gov.cn/gzdt/2013-04/10/content\_2374428.htm}.
That is to say, on average, each private car has only 0.398 parking spaces. 
If we take other types of vehicles into consideration, the situation will get 
worse with no doubt. The problem can be alleviated via many methods. But in 
this paper we focuse on the smart parking system, primarily on 
the algorithm of parking planning.
\subsection{Related Work}
T. Litman\cite{litman2013parking} presents a comprehensive implementation guide 
on parking management, including parking planning practices.
The study of M. Idris\cite{idris2009car} reviews the evolution of parking space 
occupancy detection. The detection technology makes the status of the parking space 
available for the parking management system. 
S. Zeitman\cite{zeitman1999parking} invents a communication system for the 
parking management system. They provide a method to establish communication 
between municipality, driver and parking spaces. S. Shaheen\cite{shaheen2005smart} 
documents the research and feasibility analysis for the 
design and implementation of parking management field test. Their report gives 
a description of the parking field test and its technology in detail. 
R. Lu\cite{lu2009spark} provides a solution for smart parking scheme through taking advantage of communication 
between vehicules. Our method is different since we study the problem of processing 
parking planning on a real time smart parking system, then sending the guide information to vehicles.
\subsection{Contributions}
In this paper we present an algorithm to process parking planning for a 
real time smart parking system. First, basing on some limiting conditions in 
the real world, we transform the parking planning problem which is an on-line 
problem into an off-line problem. 
Second, we establish the mathematical model by regarding this off-line problem 
as a kind of linear assignment problem.  
Third, we design an algorithm to solve this particular linear assignment problem. 
Last, we evaluate our algorithm by some simulation experiments. 
The experimental results show that our method is both timely and efficient.

\textit{Outline} The remainder of this paper is organized as follows. 
In Section \uppercase\expandafter{\romannumeral2} we describe the parking planning 
problem with mathematical model. In Section \uppercase\expandafter{\romannumeral3} we 
present the details about the method of parking planning. 
In Section \uppercase\expandafter{\romannumeral4} we present the experimental 
evaluation of the method. Finally, in Section \uppercase\expandafter{\romannumeral5} 
we conclude and outline some prospects for future work.

\section{Establishing Mathematical Model}
The smart parking system receives parking queries one by one. 
If the smart parking system processes each parking 
query immediately when it arrives, the problem to be solved is an on-line problem. 
We can solve this on-line problem easily through greedy method. For instance, 
we guide the vehicle, which is querying parking space, to the nearest available 
parking space. The experimental results presented in 
Section \uppercase\expandafter{\romannumeral5} reveal greedy method is very 
inefficient in some situations and that makes it less feasible. 

In order to make parking planning strategy be efficient in most situations, instead of processing it immediately 
we hold parking queries in a queue for a while and the number of queries we hold is a controllable parameter. By transforming 
the on-line problem into an off-line problem, we get more information and 
then we can get efficient solution in most situations. 
Let $P$ denote the set of all vehicles having parking query in the queue. 
Let $S$ denote the set of all available parking spaces included in the smart parking system. 
Let $D$ denote the set of $d_{ij}$, and $d_{ij}$ is the distance between the 
vehicle $p_i(p_i{\in}P)$ and the parking space $s_j(s_j{\in}S)$. 
We can achieve $D$ through many methods, such as GPS\cite{teunissen1998gps}. 
Let $M$ and $N$ be the size of $P$ and $S$, respectively. 
So the size of $D$ is $M{\times}N$. 

We take vehicles as jobs and parking spaces as agents. 
We take distances between vehicles and parking spaces as costs for agents 
doing jobs. We save the solution in $X$ where $x_{ij}{\in}X$. That is, 
\begin{equation}
x_{ij}=\left\{
\begin{array}{rcl}
0, &  & \text{if $p_i$ will not be guided to $s_j$;}\\
1, &  & \text{if $p_i$ will be guided to $s_j$.}
\end{array}
\right.
\end{equation}

Let $C$ be the total cost for all vehicles in $P$ going to the parking spaces 
assigned to them by the smart parking system. That is, 
\begin{equation}
C=\sum\limits_{i=1}^M\sum\limits_{j=1}^N{d_{ij}{\times}x_{ij}}.
\end{equation}

We aim to make $C$ minimum on condition that each vehicle gets exactly one parking 
space and each parking space can be assigned to only one vehicle at most. That is, 
\begin{equation}
\left\{
\begin{array}{rcl}
\sum\limits_{i=1}^M{x_{ij}}=1; & \\
\sum\limits_{j=1}^N{x_{ij}}\le 1. & 
\end{array}
\right.
\end{equation}

This optimization problem is a linear assignment problem. 

\section{Processing Parking Planning}
When $M$ is equal to $N$, the well-known Hungarian method\cite{kuhn1955hungarian}, proposed by 
H. W. Kuhn in 1955, can solve this optimization problem. 
The time complexity of that is $O(N^4)$. 
In 1971, N. Tomizawa\cite{tomizawa1971some} improved it to achieve 
an $O(N^3)$ running time. We describe the latter one as 
Algorithm \ref{alg:Hungarian} without detailed steps. 

In Algorithm \ref{alg:Hungarian}, $D_{N{\times}N}$ is the distance square matrix and 
we save the solution in $X_{N{\times}N}$.
The space complexity of Algorithm \ref{alg:Hungarian} is $O(N^2)$. 

\begin{algorithm}
\renewcommand{\algorithmicrequire}{\textbf{Input:}}
\renewcommand\algorithmicensure {\textbf{Output:} }
\caption{Hungarian Method.}
\label{alg:Hungarian}
\begin{algorithmic}[1]
\REQUIRE {the distance matrix $D_{N{\times}N}$};
\ENSURE {the solution matrix $X_{N{\times}N}$};
\renewcommand\algorithmicensure {\textbf{Initialization:} }
\ENSURE {$x_{ij}=0$ for each $x_{ij}{\in}X$};
\STATE Run the method proposed by N. Tomizawa\cite{tomizawa1971some} and save the result in $X$;
\RETURN $X$;
\end{algorithmic}
\end{algorithm}

\subsection{Available Parking Spaces are Enough}
In most situations, the available parking spaces are enough to satisfy all parking queries 
in the queue. That is to say, $M$ and $N$ satisfy $M\le N$. In fact, 
$M$ and $N$ satisfy $M{\ll}N$ in most situations. 

We can simply extend $P$ to $P'$ by adding $N-M$ virtual vehicles. 
At the same time, we extend the old distance matrix $D_{M\times N}$ 
to a new distance square matrix $D'_{N{\times}N}$ by 
setting all distances between virtual vehicles and 
parking spaces zero. If we have $d'_{ij}\in D'_{N\times N}$, that is,
\begin{equation}
d'_{ij}=\left\{
\begin{array}{rcl}
0, &  & \text{if $p_i'\in P'$ and $p_i'\not\in P$;}\\
d_{ij}, &  & \text{otherwise.}
\end{array}
\right.
\end{equation}

We can solve this new problem by running  
Algorithm \ref{alg:Hungarian} on $D'_{N\times N}$. 
H. W. Kuhn proved that the solution of this new problem includes the solution 
of the original one\cite{kuhn1955hungarian}. However, the value of $N$ is always 
very large. In practice, the value of $N$ is almost certainly in the millions. 
Because its time complexity is $O(N^3)$, the time consumed by Algorithm \ref{alg:Hungarian} 
is too long for a real time smart parking system. 

If we notice that the value of $M$ is always small and its upper bound is 
under our control, we can design an approximation algorithm to solve the original 
problem. First, we select a subset of $S$, denoted $S'$. Second, 
we extend $P$ to $P'$ which has the same size with $S'$ by adding virtual vehicles. We construct 
a new distance matrix $D'$ between $P'$ and $S'$. Finally, we run 
Algorithm \ref{alg:Hungarian} on $D'$ and convert the result into final solution. 

\subsubsection{Construct the Set $S'$}
Let $\text{SUB}_i$ be the top $M$ nearest available parking spaces for $p_i$. We construct the set $S'$ from the union of 
$\text{SUB}_i$ ($1{\le}i{\le}M$). That is, 
\begin{equation}
S'=\bigcup\limits_{i=1}^M{\text{SUB}_i}.
\end{equation}
We describe it as Algorithm \ref{alg:ConstructSubSet} in detail. 

In Algorithm \ref{alg:ConstructSubSet}, $P$ represents the set of vehicles that 
have query in the queue and $D_{M{\times}N}$ is the original distance matrix. 
We save the subset in $S'$ picked from $S$, denoting the set of all available parking spaces. 
For each vehicle in $P$, we save the top nearest $M$ available parking spaces 
in $S-Set$ temporarily. To the best of our knowledge, 
we can get the top $M$ nearest available 
parking spaces by using Heap, which is a common and useful data structure. 
The time complexity of it is $O(MN+M\text{log}_2N)$.
We check whether $s_j$ has been in $S'$ or not by using Hash method. 
The time consumed by that is constant, thus it can be omitted. 
So the the total time cost by Algorithm \ref{alg:ConstructSubSet} 
is $O(MN+M^2\text{log}_2N)$. The space complexity is $O(N+M^2)$. 

\begin{algorithm}
\renewcommand{\algorithmicrequire}{\textbf{Input:}}
\renewcommand\algorithmicensure {\textbf{Output:} }
\caption{Construct Subset $S'$.}
\label{alg:ConstructSubSet}
\begin{algorithmic}[1]
\REQUIRE {the set $P$ and the matrix $D_{M{\times}N}$};
\ENSURE {the subset $S'$};
\renewcommand\algorithmicensure {\textbf{Initialization:} }
\ENSURE {$S'={\varnothing}$};
\FOR{each vehicle $p_i{\in}P$}
\STATE get the top $M$ nearest available parking spaces \\according to $D_{M{\times}N}$ 
and save them in $S-Set$;
\FOR{each parking spaces $s_j{\in}S-Set$}
\IF{$s_j$ is not in $S'$}
\STATE add $s_j$ to $S'$;
\ENDIF
\ENDFOR
\ENDFOR
\RETURN $S'$;
\end{algorithmic}
\end{algorithm}

\subsubsection{Extend $P$ to $P'$ and Construct $D'$}
Let $N'$ be the size of $S'$ and we extend $P$ to $P'$ by adding $N'-M$ virtual 
vehicles. The distances between virtual vehicles and parking spaces in $S'$ 
are all zero. We describe the method of constructing $D'$ in 
Algorithm \ref{alg:ConstructNewMatrix}. 

In Algorithm \ref{alg:ConstructNewMatrix}, $S'$ is the subset picked from the set of all available 
parking spaces by algorithm \ref{alg:ConstructSubSet}, $D_{M{\times}N}$ is the original distance matrix. 
We save the new distance square matrix in $D'_{N'\times N'}$. We use $MAP_P$ to 
map the row index in $D'_{N'\times N'}$ to the vehicle in $P'$. 
We use $MAP_P$ to map the column index in $D'_{N'\times N'}$ to the parking space in $S'$. 
They help us to construct the new distance square matrix $D'_{N'\times N'}$. 
The time complexity of Algorithm \ref{alg:ConstructNewMatrix} 
is $O(MN')$ and the space complexity is $O(N'^2)$. 
If we notice $O(N')=O(M^2)$ and substitute it into $O(MN')$ and $O(N'^2)$, 
the time and space complexity are $O(M^3)$ and $O(M^4)$, respectively.

\begin{algorithm}
\renewcommand{\algorithmicrequire}{\textbf{Input:}}
\renewcommand\algorithmicensure {\textbf{Output:} }
\caption{Construct New Distance Matrix $D'$.}
\label{alg:ConstructNewMatrix}
\begin{algorithmic}[1]
\REQUIRE {the set $S'$ and the matrix $D_{M{\times}N}$};
\ENSURE {the index for vehicles saved in $MAP_P$, the index for parking 
spaces saved in $MAP_S$ and the new distance matrix $D'_{N'{\times}N'}$};
\renewcommand\algorithmicensure {\textbf{Initialization:} }
\ENSURE {$d'_{ij}=0$ for each $d'_{ij}{\in}D'$};
\STATE {$index=1$}
\FOR{each vehicle $p_i$}
\STATE {$MAP_P[index]=p_i$}; 
\STATE {$index=index+1$};
\ENDFOR
\STATE {$index=1$};
\FOR{each parking space $s'_j{\in}S'$}
\STATE {$MAP_S[index]=s'_j$}; 
\STATE {$index=index+1$};
\ENDFOR
\FOR{each vehicle $p_i{\in}P$}
\FOR{each parking space $s'_j{\in}S'$}
\STATE {$d'_{MAP_P[i]MAP_S[j]}=d_{ij}$};
\ENDFOR
\ENDFOR
\RETURN $MAP_P$, $MAP_S$ and $S'$;
\end{algorithmic}
\end{algorithm} 

\subsubsection{Get the Final Solution}
First, we run Algorithm \ref{alg:Hungarian} on $D'_{N'\times N'}$ and save its solution 
in $X_{N'{\times}N'}$. It costs $O(N'^3)$ running time 
and $O(N'^2)$ running space. If we notice $O(N')=O(M^2)$ and substitute it 
into $O(N'^3)$ and $O(N'^2)$, the running time and space are $O(M^6)$ and $O(M^4)$, respectively. 
Because $N'{\geq}M$ is obviously always right, the solution $X$ always exists. 
Let $X'$ be the final solution for the original problem. We can construct $X'$ from $X$ 
by the method shown in Algorithm \ref{alg:ConstructFinalSolution}. 

In Algorithm \ref{alg:ConstructFinalSolution}, $M$ is the number of vehicles 
in set $P$, $MAP_P$ and $MAP_S$ are used to map the index in $D'_{N'\times N'}$, 
$X_{N'{\times}N'}$ is the temporary solution from Algorithm \ref{alg:Hungarian}.
We save the final solution in $X'$. The time complexity of 
Algorithm \ref{alg:ConstructFinalSolution} is $O(MN')$ and the space complexity is 
$O(M)$. If we notice $O(N')=O(M^2)$ and substitute it into $O(MN')$, 
the time complexity is $O(M^3)$. 

\begin{algorithm}
\renewcommand{\algorithmicrequire}{\textbf{Input:}}
\renewcommand\algorithmicensure {\textbf{Output:} }
\caption{Construct the Final Solution $X'$.}
\label{alg:ConstructFinalSolution}
\begin{algorithmic}[1]
\REQUIRE {the size $M$ of the set $P$, the index for vehicles in $MAP_P$, 
the index for parking spaces in $MAP_S$, and the temporary 
solution $X_{N'{\times}N'}$};
\ENSURE {the final solution $X'$};
\renewcommand\algorithmicensure {\textbf{Initialization:} }
\FOR{$i$ from $1$ to $M$}
\FOR{$j$ from $1$ to $N'$}
\IF {$x_{ij}$ is $1$}
\STATE {$X'[MAP_P[i]]=MAP_S[j]$};
\ENDIF
\ENDFOR
\ENDFOR
\RETURN $X'$;
\end{algorithmic}
\end{algorithm}

The total running time and space to process parking planning in most solutions  
are $O(MN+M^2\text{log}_2N+M^6)$ and $O(M^4)$, respectively. 
Because $M$ and $N$ satisfy $M{\ll}N$ in most situations, 
the method proposed by us is much better than the naive Hungarian Method, 
whose running time and space are $O(N^3)$ and $O(N^2)$, respectively. 
In practice, the size of $S'$ denoted by $N'$ is very close to $M$ rather than $M^2$. 
This implied fact makes the method we proposed more efficient than what it looks like on the level of theory. 
We will show more details about that in Section \uppercase\expandafter{\romannumeral4}.

\subsection{Available Parking Spaces are Not Enough}
In a few situations, the relation between $M$ and $N$ is $M>N$. In other words, 
the available parking spaces are not enough to satisfy all queries in the queue. 
In order to distribute parking spaces 
fairly, we should follow the principle of \textit{first come first served}. 
So we select the top $N$ parking queries in the queue and assign the $N$ available parking spaces 
to them using Algorithm \ref{alg:Hungarian}. Because the 
value of $M$ is always small and $M$ and $N$ satisfy $M>N$, the value of $N$ is also small in this situation. 
So we can get the result from Algorithm \ref{alg:Hungarian} quickly enough in this special situation.
Moreover, we will respond the other $M-N$ parking queries with the information that there are 
no more available parking spaces. 

\subsection{Relation with Greedy Method}

We notice that our method covers the greedy method. When we set $M=1$, 
our method is exactly the greedy method. However, 
the greedy method is very inefficient in some situations.
When we set $M$ a right value, our method is efficient in most situations. 
We will see that our method is controllable and 
that point makes it more feasible than the simple greedy method. 

\section{Experimental Evaluation of the Algorithm}
We simulate different situations of parking planning in the real world by 
constructing different distance matrices $D_{M\times N}$. Because the running space is 
not a critical problem for any method, we put our attentions on the efficiency and 
running time of the algorithm. 

\subsection{Efficiency of Our Method}
The new method proposed by us is an approximation algorithm. So the total cost $C$ 
will be a little higher than the global optimal solution. We run our method on many different 
distance matrices and calculate an average. Fig.1. shows the result we have got. 
\begin{figure}[H]
\begin{center}
\includegraphics [width=91mm]{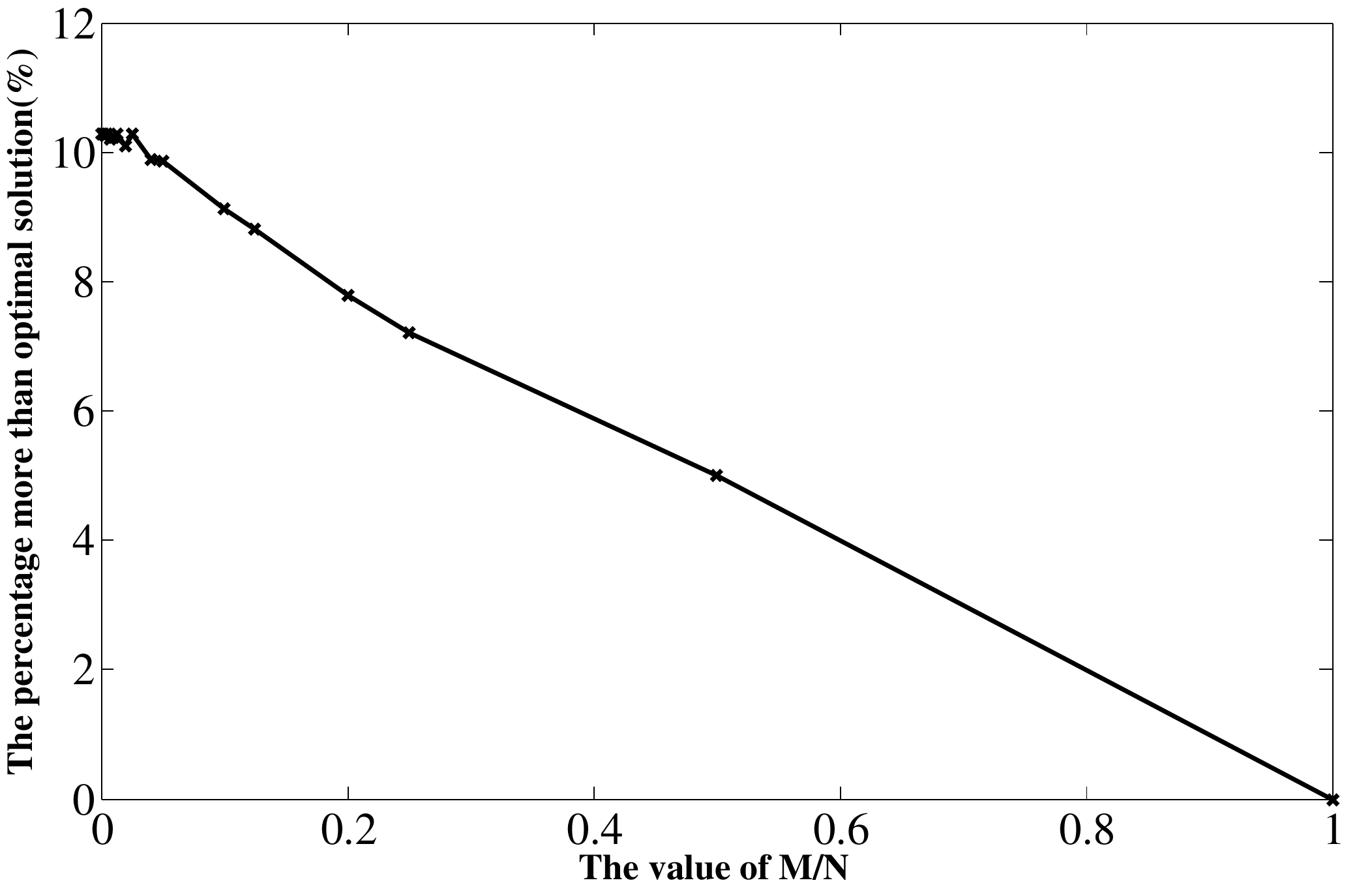}
\caption{How the efficiency of our method changes}
\end{center}
\label{fig:Accuracy}
\end{figure}

It can be seen from Fig.1. that 
the more parking queries we hold the less waste our method has. 
When $M$ and $N$ satisfy $M{\ll}N$, the waste of our method is less than 11\%. 
It is proved that our method is efficient enough for a real time smart parking 
system. 

\subsection{Running Time of Our Method}

We run our method on an old personal laptop. The CPU of it is 
\textit{Intel(R) Core(TM)2 Duo CPU P8800 @ 2.66GHz} and the memory of it 
is \textit{Kingston(R) 8.00 GB @ 1066MHz}. We set the value of $N$ as 1.6 million 
to simulate the situation of parking planning in Beijing. We run our method on many 
different distance matrices and record the average running time in
TABLE \uppercase\expandafter{\romannumeral1} in seconds. 
As we can see, our method is quick enough for a real time smart parking system. 

\begin{table}[h]
\caption{The running time of our method}
\label{f3-2}
\begin{center}
\begin{tabular}{||c|c||c|c||}
\hline
$M$ & Running Time(sec)        & $M$   & Running Time(sec) \\
\hline
\hline
1   & $1.10{\times}10^{-2}$  & 40    & $3.74{\times}10^{-1}$   \\
\hline
2   & $2.04{\times}10^{-2}$  & 50    & $4.61{\times}10^{-1}$\\
\hline
4   & $3.51{\times}10^{-2}$  & 80    & $7.46{\times}10^{-1}$   \\
\hline
5   & $5.01{\times}10^{-2}$  & 100   & $9.32{\times}10^{-1}$\\
\hline
8   & $8.24{\times}10^{-2}$  & 200   & 1.81   \\
\hline
10  & $9.02{\times}10^{-2}$  & 250   & 2.41\\
\hline
16  & $1.60{\times}10^{-1}$  & 400   & 3.90   \\
\hline
20  & $1.84{\times}10^{-1}$  & 500   & 5.14\\
\hline
25  & $2.38{\times}10^{-1}$  & 1,000 & 11.31   \\
\hline
\end{tabular}
\end{center}
\end{table}

TABLE \uppercase\expandafter{\romannumeral1} and Fig.2. also tell 
us that if we want to get results more quickly we should 
hold as few parking queries as possible.

In the real world, one parking place often has many available parking spaces. 
For instance, the number of parking spaces in each parking place in Beijing is about 3 hundred. 
As we can see from Fig.3., such clustering phenomenon makes $N'$, the size of $S'$, 
very close to $M$ and very far away from $M^2$. So the time complexity 
of our method is close to $O(MN+M\text{log}_2N+M^3)$ rather than $O(MN+M^2\text{log}_2N+M^6)$. 
This gives us the reason why the running time taken by our method 
in practice is much less than what it looks like in theory.

\begin{figure}[H]
\begin{center}
\includegraphics [width=87mm]{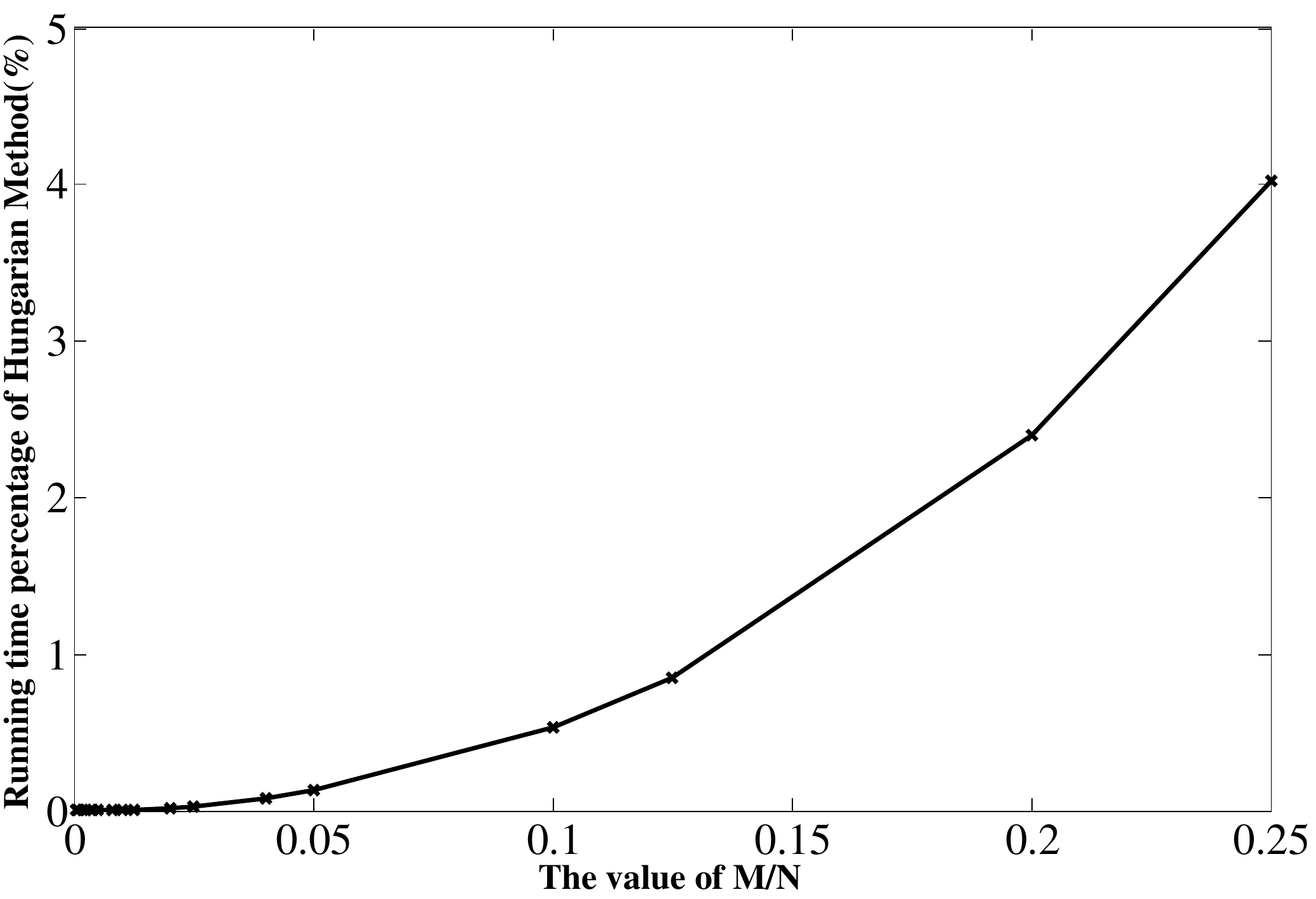}
\caption{How the running time of our method changes}
\end{center}
\label{fig:RunningTimeRatio}
\end{figure}

\begin{figure}[H]
\begin{center}
\includegraphics [width=88mm]{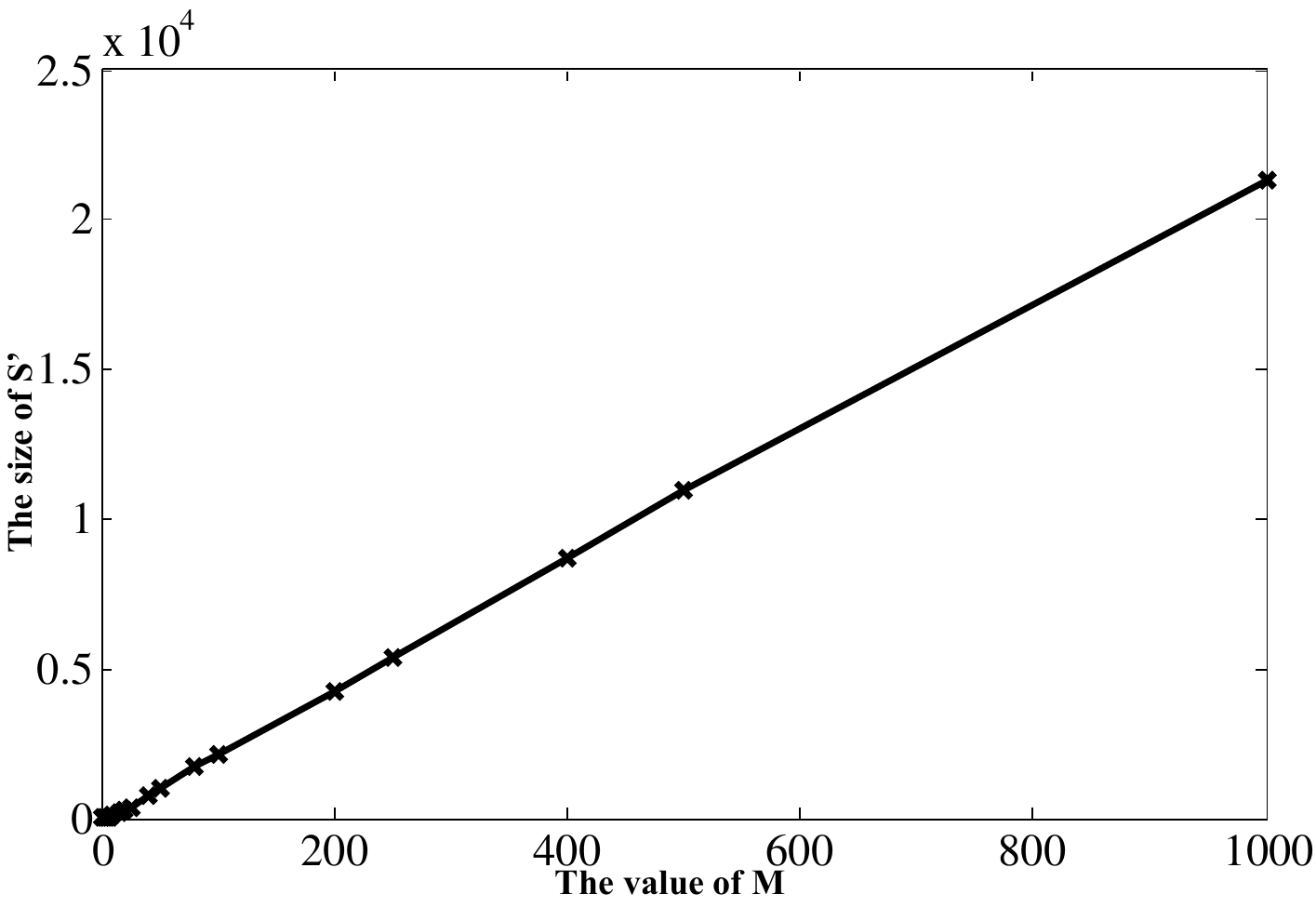}
\caption{How the size of S' changes}
\end{center}
\label{fig:SubsetSize}
\end{figure}

Above all, our method can give timely and efficient solutions 
for a real time smart parking system on condition that we choose a right value for $M$. 

\subsection{How to Choose the Best Value for $M$}
Although a small value for $M$ makes it possible for our method to get solutions quickly, 
it may also lead our method to the annoying instability. In other words, 
our method with an extremely small $M$ is very inefficient in some special situations.

To be specific, we construct a special distance matrix $D$ as an example. 
When we set $M=1$, our method is exactly the greedy method, 
the total cost of it to do parking planning is 50069. When we set $M=2$, the total 
cost  is 23549. When we set $M=3$, the total cost is 8525. When we set $M=N$, 
that is 6, the total cost is 209, which is the minimum cost in theory. 
This example helps us to illustrate that the value of $M$ should not be too small if we 
want the algorithm to be efficient in most situations. 
\begin{equation}
D=
\left( \begin{matrix}
   1 &2 &3 &4 &5 &6 \\
   1 &4 &9 &16 &25 &36 \\
   1 &8 &27 &64 &125 &216 \\
   1 &16 &81 &256 &625 &1296 \\
   1 &32 &243 &1024 &3125 &7776 \\
   1 &64 &729 &4096 &15625 &46656 \\
\end{matrix} \right)
\end{equation}

From the above, when we try to choose the best value for $M$, we should take the waste, running time and 
stability into consideration based on the real situation. 
\section{Conclusion}
We have presented a novel method of parking planning for smart parking system.
We transform parking planning into a kind of linear assignment problem by holding parking queries in a queue for a while. 
We develop a new approximation algorithm to solve this particular linear assignment 
problem. The experimental results on simulation clearly show our method is a feasible 
method which can give timely and efficient solutions for a real time smart parking 
system. 

As future work, we consider to develop methods to get an adaptive number of parking queries holden in the queue, which is denoted by $M$.  
We also consider to find ways to construct the subset of available parking spaces denoted by $S'$ more effective. For example, we can decide 
the value of $M$ according to the operation of the smart parking system and the 
historical parking data. We can try to find new ways to construct better $S'$ 
to reduce the waste and running time. All of them may also help to improve the stability. 


\bibliographystyle{IEEEtran}
\bibliography{IEEEabrv}

\end{document}